\def\year{2019}\relax
\documentclass[letterpaper]{article} 
\usepackage{aaai19}  
\usepackage{times}  
\usepackage{helvet}  
\usepackage{courier}  
\usepackage{url}  
\usepackage{graphicx}  
\usepackage{amsfonts}
\usepackage{subfigure}
\usepackage{multirow}
\usepackage{makecell}
\usepackage{booktabs}
\begin{document}
%
\title{The key to the weak-ties phenomenon}
\author{
Ke-ke Shang\textsuperscript{\rm 1,\rm 4}\thanks{Ke-ke Shang is the corresponding author},
Michael Small\textsuperscript{\rm 2,\rm 3},
Di Yin\textsuperscript{\rm 1},
Yan Wang\textsuperscript{\rm 1},
Tong-chen Li\textsuperscript{\rm 1},
\\
\textsuperscript{\rm 1} Computational Communication Collaboratory, Nanjing University, Nanjing, 210093, P.R. China\\
\textsuperscript{\rm 2} Complex Systems Group, Department of Mathematics and Statistics, The University of Western Australia, Crawley, Western Australia 6009, Australia\\
\textsuperscript{\rm 3} Mineral Resources, CSIRO, Kensington, WA, 6151, Australia\\
\textsuperscript{\rm 4} College of Big Data and Intelligent Engineering, Yangtze Normal University, Chongqing 408100, P.R. China\\
\ kekeshang@nju.edu.cn,~
\ keke.shang.1989@gmail.com
}

\maketitle
\begin{abstract}
The study of the weak-ties phenomenon has a long and well documented history, research into the 
application of this social phenomenon has recently attracted increasing attention. However, 
further exploration of the reasons behind the weak-ties phenomenon is still challenging.
Fortunately, data-driven network science provides a novel way with substantial explanatory power to analyze 
the causal mechanism behind social phenomenon. Inspired by this perspective, 
we propose an approach to further explore the driving factors behind the temporal weak-ties phenomenon.
We find that the obvious intuition underlying the weak-ties phenomenon is incorrect, and often large numbers of unknown mutual friends 
associated with these weak ties is one of the key reason for the emergence of the weak-ties phenomenon. 
\end{abstract}

\section{Introduction}
The well-known weak-ties phenomenon has a long history \cite{Granovetter1973The,Granovetter1983The} founded in social network analysis --- for example when job seekers find their new jobs they will usually rely on the persons who only have few contacts --- weak ties --- with them, in other words, the weak ties can play a more important role than strong ties in social networks. In the field of network science, Onnela et al. found that weak ties usually are the bridge links between two communities, furthermore they play the key role for the structural integrity of networks \cite{onnela2007structure,kumpula2009model}. Previous studies have also uncovered the role of bridging weak ties when modeling networks \cite{laurent2015calls}. Recent studies also have used this phenomenon to enhance the link prediction accuracy \cite{lu2010link,Shang2017}. Similar to the role of weak ties for job-hunting over time, the links with lower weights --- weak ties --- will play a more important role for the prediction of future relationships\cite{Shang2017}, that feature can also be stated as the weak-ties phenomenon. On the contrary, link prediction studies in network science also help us confirm the weak-ties phenomenon. Furthermore, we can gain substantial insight into more details for the social phenomenon via the approach of data science\cite{Deville7047,Sinatraaaf5239,hofman2017prediction,kosack2018functional,fraiberger2018quantifying,Borner12573}, and then understand the society more widely and deeply. 

Although the weak-ties phenomenon has been uncovered and applied in many recent works\cite{onnela2007structure,kumpula2009model,lu2010link,laurent2015calls} , the reasons behind the weak-ties phenomenon is still under-explored. In this paper, we adopt  network science theory -- in a specific network, where a node refers to a person, a link refers to a relationship, and the link weight refers to the relationship strength -- to explore the reasons behind the weak-ties phenomenon. Previous studies of the link prediction problem focused on the link weights itself to state or observe the weak-ties phenomenon \cite{lu2010link,Shang2017}, but these rarely paid attention to argue causality between the link weight and that phenomenon. Fortunately, Onnela et al. clearly explained the weight-topology correlations for the appear of weak-ties phenomenon \cite{onnela2007structure,kumpula2009model}.

\begin{figure}[h]

\centering

\includegraphics[width=\linewidth]{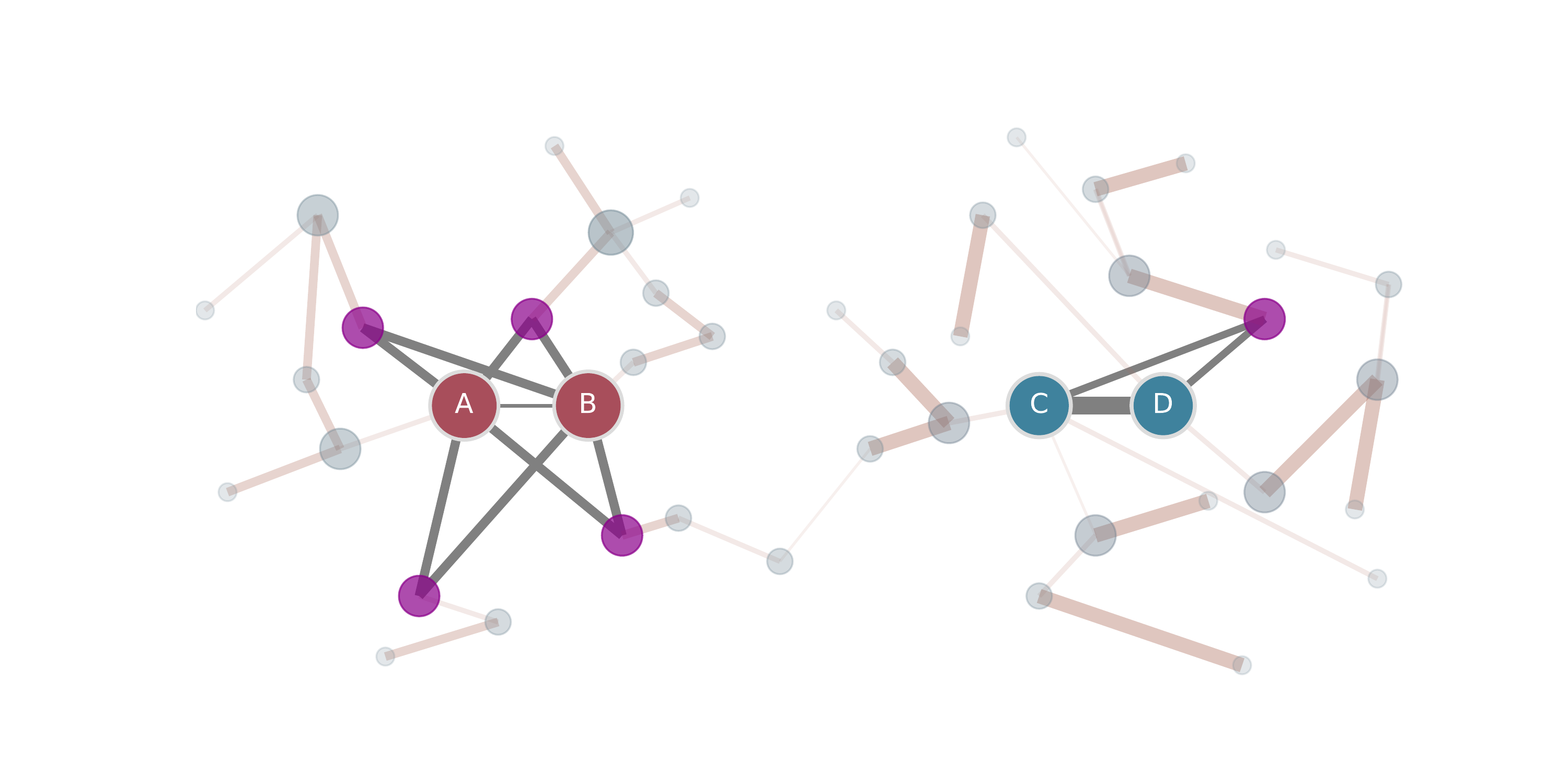}

\caption{The network structure of the weak-ties phenomenon. In the field of social science, a tie indicates the relationship between two persons, a contact or friendship can be stated as a relationship. The number of contacts can be stated as the tie strength. Ties with fewer contacts are the so-called weak ties. In the field of network science, the link or edge is precisely the tie of social science, correspondingly, link weight is the direct analogue of tie strength. Links with lower weights are weak ties. Hence, the structure of these ties or links and the properties of the persons or nodes are our primary research focus. In this figure, the node size indicates the corresponding node degree, and the link width indicates its weight. Nodes $A$ and $B$ are connected by the weak tie --- a link with low weight, and nodes $C$ and $D$ are connected by the link which has higher weight. Based on the result of our previous study \cite{Shang2017}, the pair of nodes $A$ and $B$ associated with the weak tie will have more common neighbors --- the nodes connected by nodes $A$ and $B$ at the same time, then have a higher probability of receiving a relationship between them in the future.}
\label{exam}
\end{figure}

Hence, inspired by previous studies \cite{onnela2007structure,kumpula2009model,lu2010link,laurent2015calls},  we combine the weight-topology correlations and the link prediction theory to seek the truth behind the weak-ties phenomenon.  Furthermore, we focus on an important social relationship encoded in the pair of nodes which is connected by the weak tie, that is also a measure -- the number of common friends -- of the intensity of the pair of nodes. On the other hand, in our previous study, we guessed that the number of common neighbors or friends plays an important role for the so-called weak-ties phenomenon \cite{Shang2017}. We depict our idea in Fig.\ref{exam}, if a pair of nodes have more common neighbors than that are connected by the stronger link, then they are more likely to connect to each other in the future despite the current tie or link between them is weaker. 

\section*{Network data}
As listed in Table\ref{t1}, to consider the diversity of networks and strictly compare the accuracies of different link prediction algorithms, we employ $3$ open, traditional social network datasets with different scales. 
1) Facebook \cite{Viswanath_2009}: A node represents one Facebook user, a link indicates that there is at least one Facebook "wall post'' between a pair of nodes and the number of wall posts indicates the weights between a pair of nodes. 2) Eu-core email network \cite{snapnets}: A node represents one email address of the core member from a large European research institution, a link indicates that there is at least one email between a pair of nodes, and the number of emails indicates the weights between a pair of nodes. 3) High-school contact network \cite{mastrandrea2015contact}: A node represents one high- school student, a link indicates the face-to-face contact of a pair of nodes, and the number of contact indicates the weights between a pair of nodes. 

\begin{table}[htbp]
\begin{center}
\footnotesize
\caption{The source, the number of nodes and links for $3$ networks. We do not count the number of the loop links and the duplicate nodes.}
\label{t1}
\begin{tabular*}{0.4\paperwidth}{cccc}
\hline\hline
&Facebook&Email&Contact\\ 
\hline
Nodes&$63891$&$986$&$327$\\
Links&$183412$&$16064$&$5818$\\ 
Source & Viswanath B. & SNAP Datasets & Mastrandrea R. \\ 
\hline\hline
\end{tabular*} 
\end{center}
\end{table}

\section*{Methods}

\subsection*{Link prediction problem}

Our previous study \cite{Shang2017} completely defined the link prediction problem. In this paper, we modified that description and paraphrase it as follows.
A graph $G$ can be described by a vertex set $V$, and an edge set $E$: $G=(V,E)$. Elements of the edge set are unordered pairs of elements of the vertex set: $e=(v_i,v_j)\in E$ where $v_i,v_j\in V$. The pair $(v_i,v_j)$ occurs in at most one edge $e\in E$. The standard link prediction problem, as considered by previous study \cite{lu2010link}, can be formulated as follows. The edge set $E$ is divided into two parts $E^T$ and $E^P$ where $E^T\cup E^P=E$ and $E^T\cap E^P=\emptyset$. The division into $E^P$  --- typically including $10\%$ of the observed links, and $E^T$ --- typically $90\%$ of the observed links --- is arbitrary and will be used for scoring purposes. That is, all the links in $E=E^P\cup E^T$ have been observed and are known, however, links in $E^T$ will form a {\it training set} and are used to implement a link prediction score, the efficacy of which will be evaluated over the {\it probe set} $E^P$. 

With sets $E^T$ and $E^P$, the {\it static} link prediction problem is then applied where we consider an augmentation of $G$. Suppose that the information encapsulated in the graph $G$ provides an incomplete picture of a larger
graph $G'=(V,E')$ --- the {\it truth}. This larger graph $G'$ may include both edges in $E$ and also additional edges $\bar{E}=E'\backslash E$. In the real-world these are additional edges that have not been observed.
Let $U:=\{(v_i,v_j)|v_i,v_j\in V,\;\; v_i\neq v_j\}$ be the universal set of all possible links. Links are bidirectional and hence technically $U$ should contain each link only once, so we will impose some ordering on the elements of $V$ and insist --- for example --- that $i<j$.

For each link in $U$ we define and compute a prediction measure $S_{ij}$ which measures ---  based {\it only on link information contained in $E^T$} --- how close are nodes $v_i$ and $v_j$. That is, the probability of nodes $v_i$ and $v_j$ have a link connecting each other is assessed as $S_{ij}$ using information inferred from the links in $E^T$. Using the additional links in $E^P$ we can compute a performance score for the prediction measure --- that is, based on the known links of $E^P$ or {\it known structure}--- how well correlated are the scores $S_{ij}$ with edges $(v_i,v_j,w(i,j))\in E^T$: are high values of $S_{ij}$ associated with membership of $E^P$ for edges in $U\backslash E^T$? This step ---  evaluating the score on $E^T$, is not strictly necessary, but provides a method to test how well our algorithm performs before moving to the unseen data in $\bar{E}$. Finally, our predictions of the unknown links in $\bar{E}$ can be obtained by ranking the scores $S_{ij}$ for all $(v_i,v_j,w(i,j))\in \bar{E}(U\backslash E)$. Links with highly ranked scores are those predicted to most likely exist --- we expect that these highly ranked links will probably occur in $\bar{E}$, and other few lowly ranked links will probably occur in $E^P$.

The static link prediction problem can now be stated: given $E^T$, $E^P$ and also $V$, predict $\bar{E}$ and some fake links in $E^P$. That is, if we know some of the links of a network --- those links being partitioned into the {\it training set} $E^T$ and the {\it probe set} $E^P$ --- which we have {\it observed}, is it possible to predict the existence of {\it unobserved} or {\it fake} links. The unobserved links are members of $U\backslash E$ and may be said to either {\it exist} --- they are also members of $\bar{E}$, or be {\it non-existent}, they are instead members of $U\backslash(E\cup\bar{E})$). Of course, in general, the link prediction problem is ill-posed. If the links or network structure are random and uncorrelated the information in $E$ tells us nothing about any of the remaining possible pairs $U\backslash E$ and whether they are in $\bar{E}$. However, many real-world networks exhibit correlation among the links and it is a practical problem of great importance to identify these unknown connections. In this manuscript we address two seperate questions. First, how well does an algorithm perform in predicting links? Second, how predictable is the link structure of a prediction network?

Finally, we will complicate this problem further by considering the problem of evolving
networks. In the real-world, existence of the links in the future incarnation of the network is an important problem and the way in which a network evolves over time may contain additional information about the future link structure. To state the evolving link prediction problem, we use the time point to divide
the edge set $E$ into two parts: the past edge set $E_{history}$ that includes
about 50\% of the total links weights --- these fractions are arbitrary, we provide numbers here as this is the typical treatment of these sets in the literature, and
the future edge set $E_{future}$ that includes about 50\% of the total links weights, where
$E_{history}\bigcup E_{future}=E$ and
$E_{history}\bigcap E_{future}\not\equiv \emptyset$,
$E_{history}$ and $E_{future}$ have the same vertex set $V$. A proportion of existent links in ${E_{history}}$ constitute the link set $E_{sample}$. Here, $E_{history}$ is used for computing scores, $E_{sample}$ is used for training, $E_{future}$ is used for testing, and $E_{history}$, $E_{future}$ and $E_{sample}$ are used for
the evolving link prediction accuracy measure. 

\subsection*{Metric for Link prediction in evolving networks}

$Precision$ has since been widely applied to measure the
performance of recommender systems \cite{herlocker2004evaluating} and prediction accuracy \cite{lu2010link} in a wide variety of settings. $Precision$ is defined as the ratio of relevant items selected to the number of selected items. On the other hand, our previous studies have divided the network into two parts \cite{shang2014limitation,shang2016evolving,Shang2017}: the current
network $E_{history}$ and the future network $E_{future}$. Hence,
similar to $Precision$, we propose the evolving $Precision$ ---  $P_t$ --- with a more reasonable physical significance -- to measure the prediction accuracy in evolving networks. Only the information of $E_{history}$ is allowed to be used to compute the performance score. First, to reduce the computational load, we choose a sample set
$E_{sample}$ from $E_{history}$. Here, to select a suitable sample
which has the proper density and number of nodes, we
choose a node randomly, then only select some of its neighbors and neighbors of neighbors --- up to $500$ nodes for the Facebook, and $300$ nodes for the Email and Contact, and the links between these nodes as the sample set $E$. Second, we compute the scores of all pairs of nodes from $E_{sample}$, then choose top $5\%$ or $m$ pairs of nodes by their scores. Third, if there are $n$ pairs of nodes from the top $m$ pairs of nodes will appear in the future network $E_{future}$, then the score of the evolving $Precision$ $S_{P_t} = n/m$. 

\subsection*{Algorithms for the analyzation of weak-ties phenomenon}
\subsubsection*{Link prediction Algorithms}
In what follows, we focus on studying inherent relationships among the link weights of a network, the common neighbors and the weak-ties phenomenon for furthering our arguments. We have seen previously that weaker links will play a more significant role for the link prediction \cite{lu2010link,Shang2017} both in particular for static and for evolving networks. Furthermore, the weak-ties phenomenon is counterintuitive, and has understandably received widespread scientific attention and utilization. While it is an interesting phenomenon and is even useful as a rule of thumb, it cannot be treated as axiomatic. In this contribution we seek a deeper understanding of the causes, and not just the effects, of the phenomenon. We have combined the direct link weights and the number of common neighbors to predict the future existence of links \cite{Shang2017}:
$$S_{i,j}^{DWCN}=\sum_{k\in\Gamma(i,j)} \omega(i,j)^\alpha, \eqno(1)$$
where $\Gamma(i,j)$ denotes the set of common neighbors of the
nodes $i$ and $j$, hence, $k$ denotes each common neighbor. Then, the {\it common neighbors} algorithm \cite{liben2007link} --- $S_{i,j}^{CN}=|\Gamma(i,j)|$ --- is equal to the number of common neighbors. 
$\omega(i,j)=\omega(j,i)$ denotes the weight of the direct link between $i$ and $j$. Hence, we named our algorithm as {\it direct weighted} $CN$ or $DWCN$, if the common neighbors or the weights of the pair of
nodes are 0, $S_{i,j}=0$.

Based on the fluctuation of $DWCN$ prediction performance, we have analyzed the relationship between  common neighbors and the emergence of the weak-ties phenomenon\cite{Shang2017}. To further substantiate the role of common neighbors for this phenomenon, we thus increase the effect of common neighbors, we then obtain the {\it common neighbors added} ($CNA$) algorithm:
$$S_{ij}^{CNA}=\sum_{k\in\Gamma(i,j)} \omega(i,j)^\alpha+|\Gamma(i,j)|. \eqno(2)$$
On the contrary, we also decrease the effect of common neighbors to take our argument further, then get to the {\it common neighbors decreased} ($CND$) algorithm:
$$S_{ij}^{CND}=\omega(i,j)^\alpha, \eqno(3)$$
if the weights of the pair of nodes are 0, $S_{i,j}=0$. Here, we remove the $\Gamma(i,j)$, that is to say $CND$ removes the role of common neighbors with the comparison of $DWCN$.

To this end, the previous three algorithms all consider the strength of the predicted relationship --- direct link weight --- between two persons, but with different effects of common neighbors. We can compare the $DWCN$, $CNA$ and $CND$ to asses the actual relationships among the link weight, the common neighbor and the weak-ties phenomenon. In this letter, we focus on the consequence of the basic algorithms $DWCN$, $CNA$ and $CND$, and use time -- evolving Precision $P_t$ -- to measure the prediction accuracy in evolving social networks. Under our metric, a better link algorithm is more likely to accurately predict the future link, we then earn a better score. In addition, we also propose another four new algorithms based the famous link prediction algorithms $WRA$ and $WAA$ \cite{lu2010link} to achieve a better link prediction performance for social networks.

\subsubsection*{Null models}
\begin{figure}[h]

\centering

\includegraphics[width=0.5\textwidth]{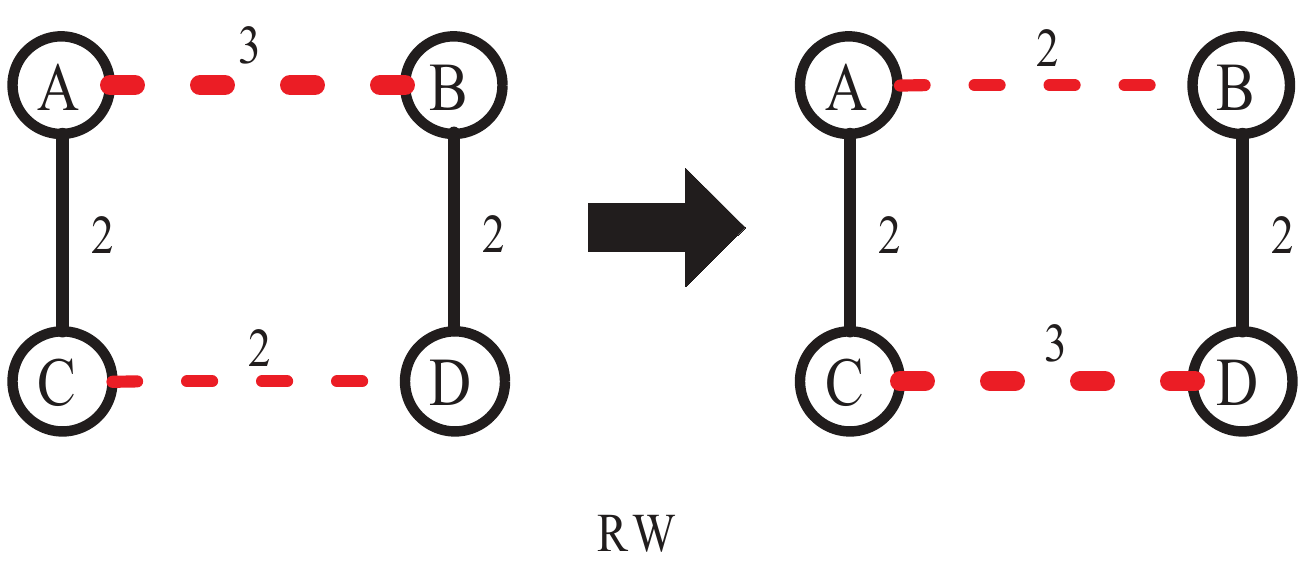}
\caption{First, we can freely choose two links that have unequal weights, in this example the weights of $AB$ are 3, and the weights of $CD$ are 2. Then we exchange the weights of the two links. Here, after the randomization, the weight of $AB$ is 2, the weight of $CD$ is 3.}
\label{fig:RW}
\end{figure}

To destroy the association between the weak ties and the common neighbors, we can use the randomized weights algorithm \cite{li2005weighted} to construct a null model for which every link weight is randomly chosen from the original network. As shown in Fig.\ref{fig:RW}, first, we can freely choose two links that have unequal weights, in this example the weights of $AB$ are $3$, and the weights of $CD$ are $2$. Then we exchange the weights of the two links. Here, after the randomization, the weight of $AB$ is $2$, the weight of $CD$ is $3$. $RW$ exchanges the weights of a pair of links.

\section*{Results}
\begin{figure*}[b]

\centering

\includegraphics[width=1\textwidth]{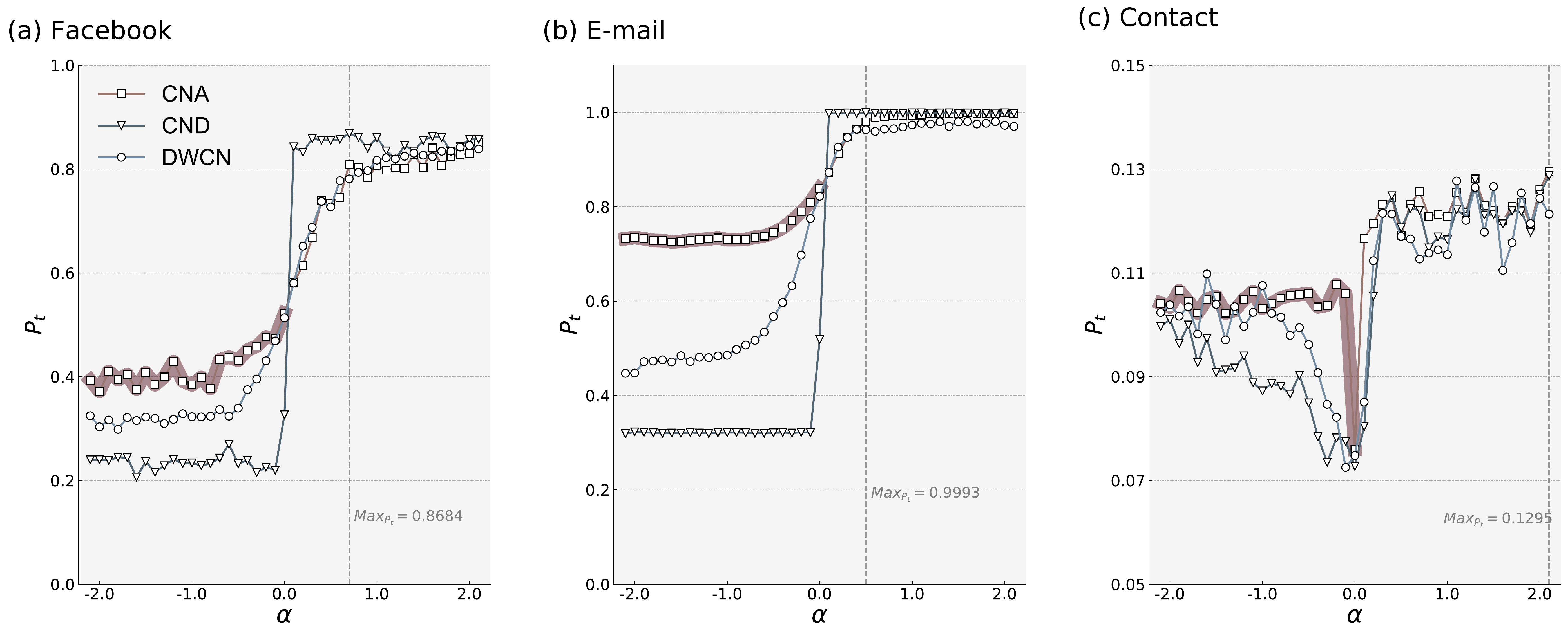}

\caption{The variation with link weight exponent $\alpha$ of the prediction accuracy $P_t$ of three link prediction algorithms, $DWCN$, $CNA$ and $CND$, applied to four different networks. Each value of $P_t$ plotted is the mean over $100$ independent trials.}
\label{r1}

\end{figure*}

For all link prediction algorithms in this paper, the negative parameter $\alpha$ means enhancing the prediction effect of lower-weight links. If the algorithm achieves the best performance when $\alpha<0$, we infer that the weak-ties phenomenon appears in the social network \cite{lu2010link,Shang2017}. 
Furthermore, the weak-ties phenomenon is coming while the algorithm performing better and better when 
$\alpha<0$, but performing very close to the initial one when $\alpha<0$. In other words, $\alpha<0$ means links with lower weights or weak ties play a more important role for the prediction. On the contrary, $\alpha>0$ means links with higher weights or strong ties play a more important role for the prediction. If the prediction performance of 
$\alpha<0$ inicreases significantly, but at the same time the prediction performance of $\alpha>0$ is similar to the initial performance, we can say that the weak-ties phenomenon is emerging, that is to say only the role of weak-ties is increasing. Hence, we can find the abnormal phenomenon: nodes with more mutual friends and weak ties are becoming more likely to connect, but nodes with more mutual friends and strong ties still maintain the original connection probability. Fig. \ref{r1} depicts the application of our method for four evolving social networks with different scales and from different domains. By shifting the attention from the direct link weight to the common neighbors, the role of common neighbors emerging from our analysis. Notably, when $\alpha<0$, for the Facebook network, contact network and email network, the $CNA$ algorithm with more common neighbor effects achieves the best performance. Conversely, the $CND$ algorithm with less common neighbor effects achieves the worst performance. However, at the same time, for Facebook network, contact network and email network, all algorithms performing similarly when 
$\alpha>0$. Hence, we can say that the weak-ties phenomenon is emerging. These phenomena suggest that the number of common neighbors with insufficient attention plays a key role for the emerging of weak-ties phenomenon. There is a strong association between large number of mutual friends and the weak ties, when the weak-ties phenomenon appearing.
These tendencies display that two persons with a weak tie prefer to communicate in the future, if they have more mutual friends. 

\begin{figure*}[h]

\centering

\includegraphics[width=1\textwidth]{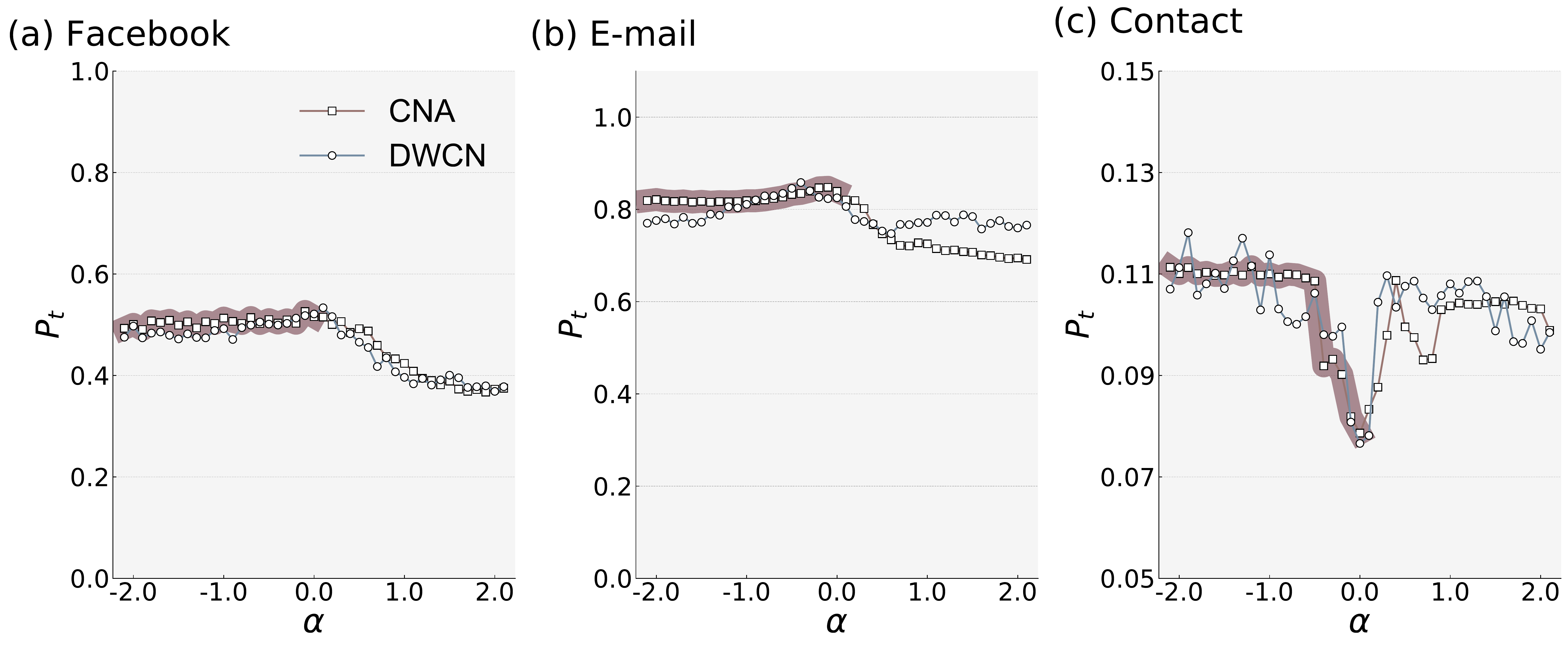}

\caption{After the randomization of link weights, the prediction accuracy of $DWCN$ and $CNA$ with the variation of $\alpha$ by the measure of $P_t$ for Email and Contact networks.}
\label{r2}
\end{figure*}

To further evaluate the reliability of the emergence of the weak-ties phenomenon \cite{holme2012temporal}, and following previous studies \cite{maslov2002specificity,shang2016evolving,Shang2017}, 
we introduce the randomized weights null model to destroy the association between common neighbors and lower-weight links. The link weight of the null model is totally random, while the network structure is maintained. The algorithm $CND$ lose its physical significance due to the randomly weights. Hence, as shown in Fig. \ref{r2}, we only observe the performance of $DWCN$ and $CNA$ for the corresponding randomized networks. As expected, the performance of the two algorithms with the common neighbors effect --- $DWCN$ and $CNA$ --- have small changes and are similar to each other in all randomized networks when $\alpha<0$. This result further demonstrates that there is a direct positive correlation between the common neighbors associated with lower-weight links and the weak-ties phenomenon. On the other hand, compared to the original networks, the emerging trend of weak-ties phenomenon is becoming more obvious in the null models, this means that the original link weights distribution plays a negative role for the emergence of weak-ties phenomenon. Our finding is due to the fact that the social pattern described in this paper is verifiable and replicable. On the other hand, without the weak ties role, when $\alpha>0$ the performance of algorithms decrease sharply due to the randomly of network structure.

\section*{Conclusion and discussion}
We show that the data-driven network approach described here is feasible for and has a deep insight into the principle of sociology, and explores the possible truth behind social phenomenon. Especially with the growth of social data information, this shift is obviously more powerful and useful. Actually, Onnela et al.\cite{onnela2007structure,kumpula2009model} have uncovered one key reason of weak ties phenomenon, some weak ties usually can play a key role for the network structure due to their special positions -- the bridges among communities. We adopt the network theory to further uncover the key reasons of the famous social phenomenon -- the mutual friends with weak ties is one of the key reason for the emergence of weak-ties phenomenon. On the other hand, our novel algorithms can achieve a better link prediction performance for the large scale networks, such as Facebook network and email network. Furthermore, null models also can test the robustness of social phenomenon via destroy the key factor step by step.


\bibliographystyle{aaai}

\section*{Acknowledgements}
Ke-ke Shang is supported by National Natural Science Foundation of China 61803047 and Tencent Research Institute. Michael Small is supported by ARC Discovery Project DP180100718.

\section*{Author contributions statement}
Ke-ke Shang conceived the experiments,  Di Yin conducted the experiments, Ke-ke Shang, Michael Small, Di Yin and Tong-chen Li analyzed the results, Ke-ke Shang, Michael Small and Yan Wang wrote the manuscript. All authors reviewed the manuscript.

\section*{Competing interests} 
The authors declare no competing interests.

\section*{Data Availability}
The datasets generated and/or analysed in this study are available from the corresponding author on reasonable request.


\section*{Supplementary information}

{ \bf Link prediction algorithms}

Our previous study (Shang et al. 2017) also proposed the direct weighted {\it Adamic-Adar algorithm (Adamic et al. 2005)} ($DWAA$): 
$$S_{ij}^{DWAA}=\sum_{k\in\Gamma(i,j)} \omega(i,j)^\alpha/ \log(1+S(k)), \eqno(1)$$
and the direct weighted {\it resource allocation algorithm (Zhou et al. 2009)} ($DWRA$): 
$$S_{ij}^{DWRA}=\sum_{k\in\Gamma(i,j)} \omega(i,j)^\alpha/S(k). \eqno(2)$$

Correspondingly, we change the above two algorithms into common neighbors added algorithms ($AAA$, $RAA$) and common neighbors decreased algorithms ($AAD$, $RAD$). 
For $AAA$: 
$$S_{ij}^{AAA}=\sum_{k\in\Gamma(i,j)} \omega(i,j)^\alpha/ \log(1+S(k))+|\Gamma(i,j)|, \eqno(3)$$
and for $AAD$: 
$$S_{ij}^{AAD}=\omega(i,j)^\alpha/ \log(1+S(k)). \eqno(4)$$
For $RAA$: 
$$S_{ij}^{RAA}=\sum_{k\in\Gamma(i,j)} \omega(i,j)^\alpha/S(k)+|\Gamma(i,j)|, \eqno(5)$$
and for $RAD$: 
$$S_{ij}^{RAD}=\omega(i,j)^\alpha/S(k), \eqno(6)$$

Here, $S(k)=\sum_{z\in\Gamma(k)} \omega(k,z)^\alpha$. If the number of common neighbors or the weights of the pair of nodes are $0$, $S_{i,j} = 0$. Table S1 provides the best performance for all algorithms.
\\
\\

\leftline{\bf Table S1}
\begin{table}[htbp]
\footnotesize
\center
\setlength{\tabcolsep}{1mm}{
\begin{tabular*}{0.43\paperwidth}{cccc}
\hline\hline
&Facebook&Email&Contact\\ 
\hline
DWCN& $0.8456$ $(\alpha=2.0)$& $0.9808$ $(\alpha=1.6)$& $0.1277$ $(\alpha=1.1)$\\
CNA& $0.8508$ $(\alpha=2.1)$& $0.9988$ $(\alpha=2)$& $0.1296$ $(\alpha=2.1)$\\ 
CND& ${\bf 0.8684}$ $(\alpha=0.7)$& $0.9993$ $(\alpha=0.5)$& $0.1287$ $(\alpha=2.1)$\\ 
DWAA& $0.8364$ $(\alpha=2.1)$& $0.9984$ $(\alpha=2.1)$& $0.1300$ $(\alpha=0.3)$\\  
AAA& $0.8253$ $(\alpha=2.1)$& ${\bf 0.9995}$ $(\alpha=1.9)$& $0.1083$ $(\alpha=1.2)$\\
AAD& $0.7335$ $(\alpha=2.1)$& $0.9985$ $(\alpha=1.9)$& $0.1255$ $(\alpha=0.8)$\\ 
DWRA& $0.8017$ $(\alpha=1.1)$& $0.9814$ $(\alpha=1.4)$& ${\bf 0.1459}$ $(\alpha=2.1)$\\ 
RAA& $0.7335$ $(\alpha=2.1)$& $0.9432$ $(\alpha=2.1)$& $0.1177$ $(\alpha=2.1)$\\  
RAD& $0.7442$ $(\alpha=1.7)$& $0.9644$ $(\alpha=1.8)$& $0.1257$ $(\alpha=-0.1)$\\  
\hline\hline
\end{tabular*}}
\end{table}

\leftline{\bf References}
Shang, K. K.; Small, M.; Xu, X. K; and Yan, W. S. The role of direct links for link prediction in evolving networks. EPL 117, 28002 (2017).

Adamic, L. A.; and Glance, N. The political blogosphere and the 2004 us election: divided they blog. In Proceedings of the 3rd international workshop on Link discovery, 36–43 (ACM, 2005).

Zhou, T.; L{\"u}, L; and Zhang, Y.-C. Predicting missing links via local information. The Eur. Phys. J. B 71, 623–630 (2009)

\end{document}